\documentclass[]{aa}
 
\usepackage{graphicx}
\usepackage{sidecap}

\newcommand{\beqa}{\begin{eqnarray}}
\newcommand{\eeqa}{\end{eqnarray}}
\newcommand{\beq}{\begin{equation}}
\newcommand{\eeq}{\end{equation}}

\newcommand{\tet}{$T_{\rm eff}$\ }

\newcommand{\lgt}{$\log g$\ }

\renewcommand{\tabcolsep}{5pt}

\begin{document}

\def\farcm{\hbox{$.\mkern-4mu^\prime$}}

\def \ms{M$_{\odot}$}
\def \rs{R$_{\odot}$}
\def \b{$\bullet$ }
\def \m{$^{\rm m}$ }
\def \lamb{$\lambda$}
\def \kms{km$\,$s$^{-1}$}
\def \D{$\Delta $}
\def \p{$\pm$}

\def \aa{A\&A }
\def \aas{A\&AS }
\def \apss{Ap\&SS }
\def \ibvs{Inf. Bull. Var. Stars No.}
\def \aj{AJ }
\def \apj{ApJ }
\def \apjs{ApJS }
\def \mn{MNRAS }
\def \pasp{PASP }
\def \pasj{PASJ }

\title{Abundances from disentangled component spectra: the eclipsing binary
       V578 Mon 
       \thanks{Based on observations obtained at the European
             Southern Observatory (ESO), La Silla, Chile}
}
\author{K. Pavlovski\inst{1} and H. Hensberge\inst{2}}

\offprints{K.~Pavlovski:~{\em pavlovski@phy.hr}}
 
\institute{Department of Physics, University of Zagreb,
 Bijeni\v{c}ka 32, 10$\,$000 Zagreb, Croatia
\and
Royal Observatory of Belgium, Ringlaan 3, B-1180 Brussel, Belgium
}

\date{Received ; Accepted }

\authorrunning{K. Pavlovski \& H. Hensberge}
\titlerunning{Abundances from disentangled spectra}

\date{Receive, Accepted}

\abstract{
Chemical abundances of the early--B type components of the 
binary V578 Mon are derived from {\em disentangled} component spectra. This is
a pilot study showing that, even with moderately high line-broadening,
metal abundances can be derived from disentangled spectra with a precision 
0.1~dex,
relative to sharp-lined single stars of the same spectral type. This binary
is well-suited for such an assessment because of its youth as a member of
the Rosette Nebula cluster NGC~2244, strengthening the expectation of an
unevolved ZAMS chemical composition. The method is of interest
to study rotational driven mixing in main-sequence stars, with
fundamental stellar parameters known with higher accuracy in (eclipsing)
binaries. The paper also includes an evaluation of the bias that
might be present in disentangled spectra.

}

\authorrunning{K. Pavlovski \& H. Hensberge}

\titlerunning{Abundances from disentangled spectra}
\maketitle

\section{\label{sec:intro}Introduction}

Stellar evolutionary models predict, besides the time-evolution
of global parameters like effective temperature, luminosity, etc.,
also changes in the chemical composition. 
Recent theoretical computations of stellar evolutionary tracks
for rotating stars suggest that the chemical
composition of the surface layers changes already when the star evolves 
on the main sequence (Meynet \& Maeder 2000, Heger \& Langer~\cite{langer}).
Observational support came the last decade from helium 
overabundances of stars in late stages of hydrogen-core burning 
(c.f.~Maeder \& Meynet 2000, and references therein).

More controversial were the results from components in close binaries. 
The first systematic study of the chemical composition of 
high-mass stars in close binaries was undertaken by Leushin~(1984).
He noticed a substantial helium enrichment for components which are still 
on the main-sequence (hereafter: MS) from an analysis of two dozen of bright 
components of binaries. His studies on carbon  (Leushin~1998a) and nitrogen 
(Leushin~1998b) disclosed carbon underabundances and nitrogen 
overabundances, as expected in the CNO bi-cycle. 
These studies had to deal with rather low-resolution, noisy photographic 
spectra and the dilution effect of the less-luminous secondary component. 
His results and these of Lyubimkov and co-workers (c.f. Lyubimkov~\cite{lyub}, 
and references therein) were contrary to the traditional view that mixing 
of the CNO bi-cycle products does not occur before  
hydrogen-shell burning. Lyubimkov~(\cite{lyub}) claimed a dependence of 
helium enrichment on the fraction of the stellar MS life-time. 
A recent review of abundance determinations using components of close binaries 
is given by Pavlovski (2004). 

The chemical analysis of binary components with 
precisely known fundamental stellar parameters 
allows a powerful comparison with 
theory. However, the precision of empirical abundances from double-lined 
binaries is hampered by increased line blending and the 
dilution of the spectral lines in the composite spectra.   
Spectral disentangling and tomographic techniques 
(Bagnuolo \& Gies 1991, Simon \& Sturm 1994, Hadrava 1995; 
c.f. Gies 2004 and Hadrava 2004 for recent reviews) overcome these 
difficulties by the separation of the individual component spectra 
using a time-series of spectra taken over the orbital cycle. 

In this paper, the first abundance analysis on disentangled B-type spectra 
is presented. It is based on the disentangled spectra obtained by 
Hensberge et al.\ (\cite{paperI}, hereafter Paper~{\sc i}) when deriving the 
orbit and the fundamental stellar parameters of the eclipsing, detached,  
double-lined binary V578\,Mon (HDE\,259\,135; NGC\,2244\,\#J8, 
Johnson~\cite{Jnrs}; NGC\,2244\,\#200, Ogura \& Ishida~\cite{OInrs}) in the 
stellar  cluster NGC\,2244 which is embedded in the Rosette Nebula. 
V578\,Mon consists of very young ($2.3\pm 0.2 \, 10^6$ yr), high mass stars, 
$M_{\rm A} = 14.54\pm0.08$~\ms\ and $M_{\rm B} = 10.29\pm0.06$~\ms. Hence,  
no abundance anomalies are expected. Moreover, the stars rotate moderately  
fast ($\simeq$~100~\kms). V578\,Mon is thus well-suited to evaluate whether  
an abundance analysis on disentangled spectra will be feasible for a large  
amount of early-B stars in close binaries.  
Following the work of Vrancken et al. (\cite{paperII}, hereafter 
Paper~{\sc ii}), the analysis is performed relative to a sharp-lined B1V star 
in the same cluster. First, the analysis method is outlined and the dominant
error sources are discussed (Sect.~\ref{sec:met}). Then, the spectral 
differences between the components are described and their disentangled 
spectra are compared to single star spectra (Sect.~\ref{sec:tg}).   
The quantitative abundance analysis is presented in Sect.~\ref{sec:ab}. 
The results are summarized 
in Sect.~\ref{sec:concl}.

\section{\label{sec:met}Method}

While the application of spectral disentangling with the 
purpose to derive orbits is common practice, the analysis of the 
component spectra is often not included. No metal abundance studies are yet 
made on disentangled spectra of OB stars, presumably because the definition
of the continuum and the progression of systematic errors to the 
disentangled spectra was not sufficiently understood. 

V578\,Mon\,A and V578\,Mon\,B rotate synchronously with the orbital  
revolution, with projected rotational velocities of 117~\kms and 94~\kms. 
A robust analysis technique is then to determine differential 
abundances relative to a template spectrum obtained by blurring the 
spectrum of a sharp-lined star of similar spectral type, as shown in 
Paper~{\sc ii}. The detailed mathematics for deriving the differential 
abundances from the comparison of lines and line blends in observed and 
model spectra is given in the appendix of that paper. It was 
shown that differential abundances derived in such a way for single 
stars rotating at more than 200~{\kms} can have an accuracy better than 
0.1~dex if the star's atmospheres are sufficiently similar.  
 
An important ingredient in the analysis was the care taken to eliminate 
systematic errors in the normalisation of the spectra with respect to the 
template. Fig.~6 in Paper~{\sc ii} shows that flux normalisation errors of low 
frequency in wavelength with an amplitude of 1\% exist, even when all spectra 
were normalised consistently using the same continuum windows. While such 
errors are small at first glance, they are not negligible relative to 
the depth of the relevant spectral lines in rotating early-B stars. 
Systematic normalisation errors were estimated and eliminated by applying the 
normalisation procedure to the (not yet normalised) artificially broadened 
sharp-lined star and comparing that with the spectrum obtained by broadening 
the normalised sharp-lined spectrum. Fig.~7 in Paper~{\sc ii} shows that the 
spectra of fast rotating stars of the same spectral type are then within the 
noise identical to the broadened template taking into account the effects of 
a small difference in temperature.

\subsection{\label{sub:data}Data}

For the analysis of V578\,Mon, we use the same template star as in 
Paper~{\sc ii}, NGC\,2244\,\#201 (B1V, $v \sin i = 22$ {\kms}). 
It has the same spectral type as V578\,Mon\,B 
and is less than 1\farcm5 away from V578\,Mon, in the SE part of   
the cavity in the Rosette Nebula. A hotter sharp-lined member,   
NGC\,2244\,\#180 (O9.5V, $v \sin i = 24$ {\kms}) is   
still a better match to V578\,Mon\,A. \#180 has  
been recently analysed in detail by Daflon et al.\ (\cite{ab180}), who 
derive \tet = 31\,500 K and \lgt = 4.2, from high-dispersion spectra  
taken with the {\sc FEROS} echelle spectrograph at the 1.5\,m  
telescope of ESO at La Silla. From Walraven photometry, 
Verschueren~(\cite{PhD}) estimated earlier \tet = 32\,060~K and \lgt = 4.25. 
Hence, V578\,Mon\,A (\tet = 30\,000~K) is well bracketed in between the two 
comparison stars.  
The template spectra were obtained with {\sc CASPEC} at the 3.6\,m telescope 
of ESO, as the  V578\,Mon spectra, 
and reduced in the same way. Details of observations and data reduction 
are mentioned in Paper~{\sc i}, Paper~{\sc ii} and references therein. 
The impact of data reduction errors on disentangled spectra is discussed 
in Hensberge (\cite{herman}). 

\subsection{\label{sub:err}Error progression}

The observed, composite spectra reflect the characteristics of two different
stars and the lines, roughly twice as numerous and twice less deep, are
Doppler-shifted according to the orbital movement. Hence, continuum windows
are scarcer than in single star spectra and are phase-dependent. 
Although this did not prevent to normalise all the composite spectra in a 
consistent way (Sect.~4.1 in Paper~{\sc i}), application of the method used 
in Paper~{\sc ii}, to eliminate differential normalisation errors between the
spectra of the binary and the template is not evident. Both the facts that 
the spectra of V578~Mon are double-lined and that spectral disentangling was 
applied are sources of systematic normalisation errors which need 
consideration. 

The disentangling process determines in general the contribution 
$ s_{\rm A,B} $ of each component to the observed spectra up to additive 
constants $ c_{\rm A} = - c_{\rm B} $. 
 The latter indeterminacy follows from the fact that pure continuum cannot be  
uniquely disentangled, as it contains no time-dependent Doppler information. 
The observed composite spectra $O(\lambda; \Delta\lambda_{\rm A}, \Delta\lambda_{\rm B})$ 
are recovered as
\beq 
O(\lambda; \Delta\lambda_{\rm A}, \Delta\lambda_{\rm B}) = 
s_{\rm A}(\lambda; \Delta\lambda_{\rm A}) + 
s_{\rm B}(\lambda; \Delta\lambda_{\rm B}) + c_{\rm obs} 
\eeq
where the average value of $s_{\rm A}$ and $s_{\rm B}$ are zero,  
$c_{\rm obs}$ is the average value of $O(\lambda)$ over the considered 
wavelength range, $1 - c_{\rm obs}$ being the line blocking coefficient
$b_{\rm obs}$  over 
the considered wavelength range and $\Delta\lambda_{\rm A,B}$ are the 
time-dependent orbital Doppler shifts. 
While the disentangling gives the shape of the spectral features in  
$s_{\rm A}(\lambda)$ and $s_{\rm B}(\lambda)$, the sum of   
$s_{\rm A}(\lambda) + 0.5 c_{\rm obs} + c_{\rm A}$ and 
$s_{\rm B}(\lambda) + 0.5 c_{\rm obs} - c_{\rm A}$ reproduces the  
composite spectra irrespective of the value of $c_{\rm A}$. 

Only when composite spectra are available in which the components contribute 
with time-dependent light fractions $ \ell_{\rm A,B} $, i.e. 
\begin{eqnarray}
\lefteqn{O(\lambda; \Delta\lambda_{\rm A}, \Delta\lambda_{\rm B}, t) = }
\nonumber \\
 &  &  \;\;\;\;\;\;\;\;\;\;\; \ell_{\rm A}(t) S_{\rm A}(\lambda;
\Delta\lambda_{\rm A})
+ \ell_{\rm B}(t) S_{\rm B}(\lambda; \Delta\lambda_{\rm B})
\end{eqnarray}
the disentangling process can recover uniquely the component spectra.  
Here $ S_{\rm A,B} $ are the component spectra normalised to their intrinsic 
continuum, and $ \ell(t) $ are relative light fractions in the continuum (with 
a slow dependence on wavelength) such that 
$ \ell_{\rm A}(t) + \ell_{\rm B}(t) = 1 $.  
As an example, a spectrum in total eclipse reveals straightforwardly one of 
the component spectra. But any factor that makes the depth of the spectral 
features of one of the components substantially time-variable in the 
composite spectra suffices to remove the indeterminacy discussed above. 
In that case, the $ \ell(t) $ are treated either as free parameters in the 
disentangling process, or specified in the input when information from light 
curves is considered more robust than adding free parameters. The latter 
applied for the eclipsing binary V578~Mon. 

For time-independent $ \ell $, the zero-points and scaling of the  
$ S_{\rm A,B} $ are not determined uniquely. Iliji{\'c} et al.\ 
(\cite{dubnor}) show how the zero-points of the normalised, intrinsic
component spectra are coupled by the observed line blocking in the composite 
spectra. 
 In the notation introduced here, this coupling may be expresses as
\beq
S_{\rm A,B} = 1 - b_{\rm obs} + \frac{s_{\rm A,B} + C_{\rm A,B}}{\ell_{\rm A,B}}
\eeq
with
\beq
C_{\rm A} = - C_{\rm B} = 
  { {b_{\rm obs} \ell_{\rm A} ( 1 - {b_{\rm A}\over b_{\rm B}} )}
\over { 1 +  {\frac {b_{\rm A}}{ b_{\rm B}} {\ell_{\rm A} \over \ell_{\rm B}}}  }}
\eeq
where $ b_{\rm A} $ and $ b_{\rm B} $ are the line blocking
coefficients in the intrinsic component spectra.
Without additional astrophysical constraints,
they are only coupled by the requirement to reproduce the observed
line blocking,
\beq
\ell_{\rm A} b_{\rm A}  +  \ell_{\rm B} b_{\rm B} = b_{\rm obs}
\eeq
such that one of them (or e.g. their ratio) can be chosen freely
in the case of time-independent light fractions.
Astrophysical constraints may apply in some cases.
An evident condition is that the flux in the deepest
absorption lines should be non-negative.
This constraint is useful for late spectral type components
(see e.g. Griffin 2002) where e.g. the core of the
Ca~{\sc ii}~K line is very deep, quite independent of
temperature, but not in our specific case.

The key point to our discussion here is that the $ S_{\rm A,B} $ are obtained 
through a linear transformation of the $s_{\rm A,B}$ that involves as well 
a {\em multiplicative} as an {\em additive} operation. 
The multiplicative operation amplifies random and systematic errors by a 
factor inversely proportional to  $ \ell_{\rm A,B} $ i.e. a factor 1.45 
for V578~Mon~A and 3.2 for V578~Mon~B. This amplification allows to recognize 
low-frequency systematic errors best in the normalised spectra of the fainter 
component. For random errors, this amplification is counteracted by the use 
of several observed spectra. With 12 input spectra with a typical S/N-ratio 
of 125--250 in the 4000-4700~{\AA} wavelength region, the S/N-ratio in the 
normalised component spectra is about 360 for V578~Mon~A and 160 for 
V578~Mon~B. 

\begin{figure}[th!]
\includegraphics[width=9.2cm]{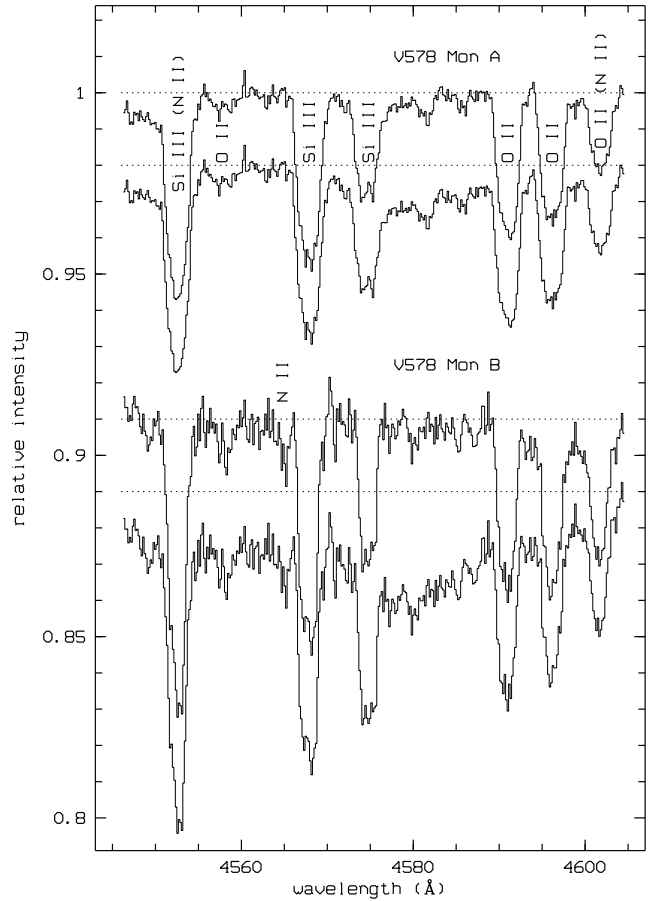}
\caption{Extract of the disentangled component spectra around a blemish in
the input spectra. The original result is shifted down by 0.02 relative to
the one obtained from improved input spectra. The spectra of V578~Mon~B are
shifted down 0.09 relative to V578~Mon~A.
}
\label{fig:dub}
\end{figure}

The difference in the additive terms reflects the difference in the line 
blocking for the two components. Hence, from the viewpoint of an abundance 
analysis, an error in these coupled additive terms would bias the abundances 
of one component in the opposite sense than for the other. 
With well-determined $ \ell_{\rm A,B} $, from photometry (the average line 
blocking is extremely similar for the components of V578~Mon, such that only 
small corrections apply to derive the light ratio in the continuum), the  
additive terms are not a dominant error source. 

Turning toward the effect of small phase-dependent normalisation errors in 
the composite spectra, we note that they may transform after Fourier space 
disentangling into undulations in the continuum of the disentangled spectra 
(Iliji{\'c} \cite{dubsys}). The disentangled V578~Mon~B spectrum is 2.2 times 
more sensitive to these than the V578~Mon~A spectrum. Inspection of the 
component spectra showed a posteriori marginal evidence for low frequency 
errors, at a level of 1\%, in the continua of the composite spectra relative 
to the continuum definition of the template star \#201. We eliminated these,
 by straightening each disentangled region in $ S_{\rm A,B} $ by a low-amplitude 
sinusoidal function, properly scaled and phased for $ S_{\rm B} $ relative to 
$ S_{\rm A} $ such that the predicted composite spectrum is not affected. This 
procedure is subjective. Independent trials suggest that spurious 
low-frequency patterns have been removed  
up to the level of 1\% in the intrinsic V578~Mon~B 
spectrum and better in regions with weaker metal lines. 

Finally, we mention that the presence of an unrecognised third component 
in the composite spectra, be it of astrophysical origin (another star, or 
e.g.\ diffuse interstellar bands or telluric lines) or of instrumental 
origin (due to detector blemishes) may lead to complex disturbances. 
We removed the interstellar CH~$\lambda$4233 line from the observed spectra 
before disentangling and we did not disentangle the spectra around the diffuse 
interstellar bands (DIBs) at $\lambda$4430 and $\lambda$4505. In the 
$\lambda$4565-4610 region, a broad absorption feature centered at 
$\lambda$4580 with a central depth of 1.2\% in the observed spectra may be 
present; it ``grows'' in the intrinsic V578~Mon~B spectrum to almost 4\% which 
makes it easy recognisable. The feature discussed here appears too far off in 
wavelength to attribute it to the 27.6~{\AA} wide (fwhm) DIB at $\lambda$4594 
(Jenniskens \& D\'{e}sert \cite{dib94}, Tuairisg et al.\ 2000).
We do not exclude that such shallow feature is the result of a 
non-linear reacting detector column crossing the extraction slit of the 
involved echelle order. Since this ``blemish'' affects the region of the 
Si~{\sc iii} triplet at $\lambda\lambda$4553,~4568,~4575 {\AA} 
 and the O~{\sc ii} line at $\lambda\lambda$4590,~4596 {\AA}, this feature 
was removed from the observed spectra requiring that no spurious broad feature 
should appear in the disentangled spectra (Fig.~\ref{fig:dub}).
Re-defining the continuum on the observed spectra is more consistent than the 
alternative to define locally a pseudo-continuum for the equivalent width 
measurements on the affected Si~{\sc iii} and O~{\sc ii} lines in the biased 
component spectra (shown in Fig.~\ref{fig:dub}, down-shifted spectrum A and 
B). It preserves in a natural way the coupling between the corrections 
in the two component spectra. Fig.~\ref{fig:dub} suggest that the remaining 
bias in the continuum is an order of magnitude smaller than in the original 
component spectra.

\begin{figure*}
\includegraphics[width=12cm]{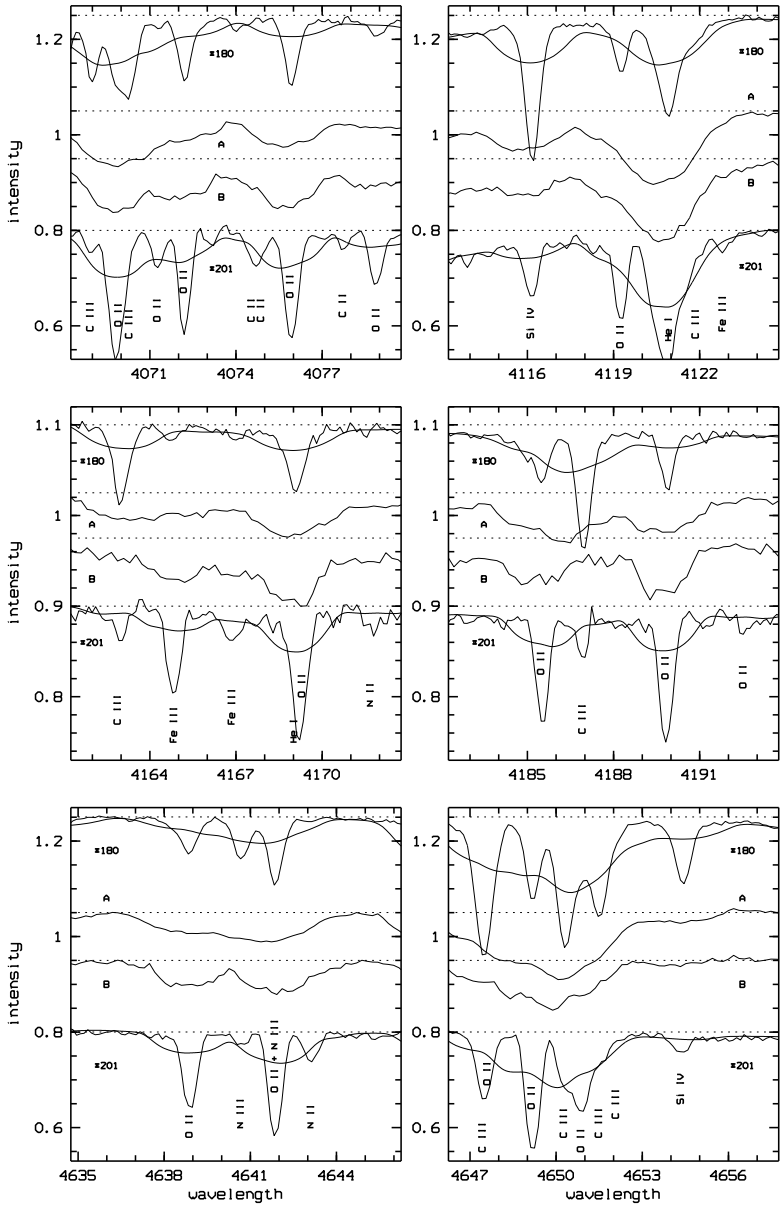}
\caption{Six spectral regions containing temperature-dependent lines and
line blends. Each panel shows, from top to bottom: the sharp-lined O9.5 \#180
and an artificially blurred version mimicking the rotational broadening in
V578~Mon~A; the spectra of V578~Mon~A and V578~Mon~B; and the sharp-lined B1V
\#201 and an artificially blurred version mimicking the rotational broadening
in V578~Mon~B. The most important absorbers are identified. The applied
vertical shift to the spectra is indicated by dashed lines representing the
continuum levels.}
\label{fig:six}
\end{figure*}

\begin{table*}
\caption{\label{tab:lis} Spectral features used in the  abundance analysis. 
Main contributing ions are indicated. For each binary component, the  
equivalent width $W$ of all the absorption over the spectral interval 
specified in columns 2 and 4 is given in columns 3 and 5. The values 
between brackets are the corresponding $W$ measured in the \#201 spectrum 
blurred to the rotation velocity of component A resp. B (and in \#180, 
behind the semicolon in the fourth column). }
\begin{tabular}{|l|cr@{ }r@{ ; }r|cr@{ }r|}
\noalign{\smallskip}\hline\noalign{\smallskip}
contributing ions & $\Delta\lambda_{\rm A}$ & 
\multicolumn{3}{l|}{$W_{\rm A}$} & 
$\Delta\lambda_{\rm B}$ & \multicolumn{2}{l|}{$W_{\rm B}$} \\
 & \AA  & \multicolumn{3}{l|}{m\AA}  & \AA  & \multicolumn{2}{l|}{m\AA}  \\
\noalign{\smallskip}\hline\noalign{\smallskip}
O\,{\sc ii} + C\,{\sc iii} & 4066.31--4073.90 & 427 & (325 & 433) & 4067.14--4073.60 & 277 & (324) \\  
O\,{\sc ii}     & 4073.90--4080.71 & 261 & (239 & 183) & 4073.60--4080.71 & 248 & (245)  \\
O\,{\sc ii} (C\,{\sc iii}) &  & $-$ & ($-$ & $-$) & 4188.13--4191.32 &  114 & (100) \\
Si\,{\sc iv}    & 4210.25--4214.50 & 40 & ($<$16 & 63)   &  & $-$ &($-$) \\
C\,{\sc ii}     & 4264.80--4269.20 & 108 & (122 & 59) & 4265.60--4269.20 & 139 & (123) \\ 
O\,{\sc ii}     & 4273.70--4279.65 & 199 & (213 & 118) & 4273.70--4279.23 & 213 & (211) \\ 
Mg\,{\sc ii} + Al\,{\sc iii} & 4479.15--4483.30 & 121 & (143 & 136) & 4479.66--4482.40 & 140 & (129) \\ 
Si\,{\sc iii} + N\,{\sc ii} & 4550.13--4555.15 & 162 & (159 & 101) & 4550.13--4554.54 & 167 & (156) \\ 
Si\,{\sc iii}   & 4565.60--4570.05 & 135 & (153 & 98) & 4566.07--4569.76 & 137 & (150) \\ 
Si\,{\sc iii}   & 4572.55--4577.05 & 87 & (99 & 48$^1$) & 4572.70--4576.97 & 83 & (99) \\ 
O\,{\sc ii} (N\,{\sc iii}) & 4588.70--4593.10 & 104 & (117 & 84) & 4589.07--4592.64 & 100 & (113) \\ 
O\,{\sc ii}     & 4593.80--4598.56 & 105 & (115 & 50) & 4594.32--4598.56 & 117 & (112) \\ 
O\,{\sc ii} + N\,{\sc ii} (N\,{\sc iii}) & 4605.00--4611.85 & 98 & (101 & 60) & 4605.35--4611.67 & 112 & (99) \\ 
O\,{\sc ii} + N\,{\sc ii} + N\,{\sc iii} & 4636.35--4644.95 & 291 & (251 & 261) & 4636.50--4644.86 & 291 & (252) \\ 
O\,{\sc ii} + C\,{\sc iii} + Si\,{\sc iv}$^2$ & 4645.05--4656.40 & 653 & (475 & 871) & 4644.86--4652.90 & 379 & (438) \\ 
O\,{\sc ii}     & 4659.35--4663.60 & 103 & (99 & 89$^3$) & 4659.75--4663.28 & 93 & (97) \\
\noalign{\smallskip}\hline
\end{tabular} \\
Notes: \\
$^1$ with contribution from weak, unidentified line at $\lambda$4573.2 \\
$^2$ only in the $\Delta\lambda_{\rm A}$ interval  \\
$^3$ weak C\,{\sc iii} lines at edges of interval contribute partly
\end{table*}

\section{\label{sec:tg}Component Spectra}

In order to evaluate to which level the intrinsic component spectra can be 
interpreted, a detailed comparison was made with the sharp-lined stars 
\#201 and \#180. We compared blended features whose shape changes with 
temperature as a consequence of the behaviour of its constituents in the 
B1V--O9.5V spectral range. Six examples in Fig.~\ref{fig:six} illustrate that 
the disentangling process reproduces successfully the temperature-dependent 
morphology of the broad line blends: 

- $\lambda$4169 is a blend of He~{\sc i} $\lambda$4169.0 and O~{\sc ii} 
$\lambda$4169.28 whose effective wavelength shifts to the blue with higher
temperature. Nearby, the C~{\sc iii} line at $\lambda$4162.86 and the 
line at $\lambda$4164.8 form a blend whose morphology depends highly on 
temperature. 

- $\lambda$4186 is a blend of O~{\sc ii} $\lambda$4185.46 and C~{\sc iii} 
$\lambda$4186.90 with the oxygen line weakening with increasing temperature 
and the carbon line strengthening. 

- $\lambda$4641 is a blend of O~{\sc ii} lines, N~{\sc ii} and N~{\sc iii},
with N~{\sc iii} $\lambda$4640.64 filling in the central dip in the blend 
at higher temperature, and N~{\sc ii} $\lambda$4643.09 extending the 
blend to longer wavelengths at lower temperature. 

- the strength of Si~{\sc iv} $\lambda$4116.10 relative to He~{\sc i} 
$\lambda$4120.9 (the latter blended with several O~{\sc ii} lines) is very 
sensitive to temperature. 

- complex blends of, mainly, O~{\sc ii} and C~{\sc iii} are seen in the blue
wing of H~$\delta$ and near $\lambda$4650, producing wide features whose 
morphology and/or position of maximum absorption changes with temperature.

Hence, the blended line profiles in the disentangled spectra contain  
trustworthy information and the main limitation in interpreting 
disentangled spectra lies in the occurrence of low-level low-frequency 
patterns as discussed in Sect.~\ref{sub:data}.   
 As in the case of single stars, line blending due to rotational broadening 
remains a factor limiting the information on abundances in complex blends. 

We now describe shortly the aspects in which the intrinsic component 
spectra differ. The Balmer {\em hydrogen} lines and the {\em helium} lines 
have been discussed extensively in Paper~{\sc i}, Sect.~6.3.2/3 and Figs. 8--9. 
We pointed out that the hydrogen lines of V578~Mon~A are slightly stronger 
than expected considering the dynamically derived gravity and the 
temperature-sensitive He and Si lines from different ionisation stages. 
The He~{\sc i} lines are slightly stronger in the spectrum of V578~Mon~B
than in V578~Mon~A, but the difference is somewhat less than expected. 
The spectroscopic temperatures of 30\,000 $\pm$ 500~K and 26\,400 $\pm$ 400~K 
are a compromise.       
{\em Carbon} is present in two ionisation stages. Numerous C~{\sc iii} lines 
contribute to blends in which O~{\sc ii} is generally the main contributor. 
C~{\sc ii} $\lambda$4267 is prominent and unblended. The C~{\sc ii} line is 
significantly stronger in the cooler component. 
{\em Nitrogen} is present in two ionisation stages, but N~{\sc ii} is the
dominant ion in V578~Mon~B while N~{\sc iii} is dominant in V578~Mon~A. This 
is seen very clearly in the feature just longward of $\lambda$4630. In the 
cooler component N~{\sc ii} $\lambda$4630.55 dominates over 
Si~{\sc iv} $\lambda$4631.24 and N~{\sc iii} $\lambda$4634.16, but the latter
contributors shift the feature to longer wavelength in the hotter component. 
Another example is the complex blend near $\lambda$4640 shown in  
(Fig.~\ref{fig:six}). The fact that N~{\sc ii} still contributes in the 
hotter component is obvious around $\lambda$4040, where the absorption in
both stars is dominated by four lines of ionized nitrogen 
($\lambda$$\lambda$4035.08, 4041.31, 4043.53, 4044.78), around $\lambda$4240 
($\lambda\lambda$4237.0, 4241.79) and in the strong, relatively isolated 
line N~{\sc ii} $\lambda$3995.00. Many lines of ionized {\em oxygen} are 
seen at these temperatures, and several are unblended or the main contributor 
to slightly blended features, even at these appreciable rotation velocities. 
Lines of ionized {\em neon} are too weak to allow a meaningful analysis, but 
are suspected to give a minor contribution to some features. 
The doublet of ionized {\em magnesium} at $\lambda$4481 is prominent and 
weakly blended with lines of Al~{\sc iii} and O~{\sc ii}. 
{\em Aluminium} contributes to some features, mainly Al~{\sc iii} 
$\lambda$4529.19 in a blend with N~{\sc ii} $\lambda$4530.41 and, in the 
hotter component, with N~{\sc iii} $\lambda\lambda$4527.89, 4530.86.
{\em Silicon} has prominent lines from two ionisation stages, namely the 
triplet of Si~{\sc iii} at $\lambda\lambda$4552.616, 4567.82, 4574.76 
and the Si~{\sc iv} lines at $\lambda$4088.85, 4116.10. More Si~{\sc iv} lines 
are visible, e.g. at $\lambda$4212.41 and $\lambda$4631.24 (at least in 
V578~Mon~A) and $\lambda$4654.32. {\em Sulphur} and {\em iron} contribute 
at these rotation velocities at best in a subtle way to blends.

\section{\label{sec:ab}Quantitative Abundance Analysis} 

LTE line-blanketed atmosphere models were calculated for a relevant grid of 
effective temperatures and gravities, in steps of 1\,000\,K and 0.1 dex 
respectively, for solar composition and a depth-independent microturbulent 
velocity of 2~{\kms}, with the {\sc Atlas$\,$9} code (Kurucz~\cite{atlas}). 
Non-LTE line formation 
and spectrum synthesis computations were than performed as explained in 
Paper~{\sc ii} using codes {\sc detail} and {\sc surface} developed and
maintained by Keith Butler. 
The Mg~{\sc ii} atom model is due to Przybilla et al.\ (\cite{mgII}).  
 The use of the same models as in Paper~{\sc ii} allows a consistent 
differential analysis. 

The spectral features used in the quantitative analysis are listed in 
Table~\ref{tab:lis}. 
Differential abundances relative to \#201 are listed in Table~\ref{tab:abu}
and converted to relative-to-solar values using the results of 
Paper~{\sc ii} for the template star. Solar abundances are as in Daflon et 
al.\ (\cite{ab2004}). None of the V578~Mon abundances differs significantly 
from those of the template star. The tendency for $\Delta \epsilon_A$ and 
$\Delta \epsilon_B$ to have opposite signs (4 cases out of 5) might contain a 
hint of coupled normalisation errors as discussed in Sect.~\ref{sub:err}, but 
has low statistical significance. The consistency of the results rather shows 
that disentangled component spectra can be normalised sufficiently well to 
obtain abundances with an accuracy of 0.1 dex. 

With \#201, \#80, and \#128 analyzed by Vrancken et al.\ (1997) and the
components of V578~Mon, we have five early-B type stars in NGC~2244
($T_{\rm eff}$ = 26$\,$300 -- 30$\,$000 K) for which abundances are derived in a
consistent way. 
They include slow and fast rotators ($v \sin i$ = 22 -- 260 \kms). The
abundances averaged over these five stars and the corresponding rms
values are listed in Table~\ref{tab:abu2}. Within the achieved accuracy, the
stars have an identical chemical composition. No correlation with projected
rotation velocity is noticed. In passing, we note that evidence of an 
anomalous abundance pattern in the atmosphere of the young stars of NGC~2244 
is found in the strongly magnetic, less massive mid-B star \#334
(Bagnulo et al.\ \cite{334}) and in \#180. \#334  shows an abundance pattern 
typical of chemical peculiar stars of the He-weak Si subclass. Some stars may
thus modify their atmospheric chemical composition early in their youth, 
probably even before arriving on the main sequence. \#180 has been found by 
Daflon et al.\ (\cite{ab180})  to have lower metallicity. It is actually the 
star that deviates most from the average relation of metallicity versus 
galactocentric distance (Daflon \& Cunha \cite{grad}). Unfortunately, it is 
the only member of NGC~2244 in their sample. The reason for this discrepancy 
deserves attention (Daflon et al.\ (\cite{ab180}) derive e.g. a much higher 
microturbulent velocity than for the stars our group analysed), but this falls 
outside the scope of this paper. We used \#180 only to visualise the 
dependence of the morphology of temperature-sensitive line blends, but did 
not use the star in the quantitative abundance analysis in view of the 
possible break-down of the LTE line-blanketed atmosphere models at higher 
temperatures. At present we conclude that there is no compelling evidence 
against the hypothesis that the young early-B stars of NGC~2244 have all a 
chemical composition similar to the stars in the large inner-disk sample of 
Daflon et al.\ (\cite{ab2004}, see Table~\ref{tab:abu2}).

 No study of the chemical abundances in the Rosette Nebula using  
H~{\sc ii} region recombination or collisionally excited emission lines 
has been undertaken so far. Using the results on several other H~{\sc ii} 
regions and the relation with galacto-centric distance derived by Esteban et 
al.\ (2005) suggests [C/H] = 8.43 and [O/H] = 8.63, similar to modern solar 
abundances. The abundances of the early-B stars in NGC~2244, as well as 
those of early-B stars in general, are slightly lower than the solar 
abundances and those derived from H~{\sc ii} regions. Daflon \& Cunha (\cite 
{grad}) find agreement between abundances derived in massive stars and the 
accompanying H~{\sc ii} region within 0.1 to 0.2 dex. With the recent lesson 
learned from the sensitivity of solar abundances to sophisticated 
atmosphere modeling, one should probably not overinterprete difference on 
this level between abundance determinations derived with very different 
techniques. It may indicate limitations in one or several of the techniques  
rather than being astrophysically significant.

\begin{table}
\caption{\label{tab:abu} Abundances for both components of V578\,Mon, 
$\Delta \epsilon$ relative to \#201, and $[X/H]$ relative to solar.
Solar abundances are as listed in Table 3. }
\begin{tabular}{crrrrrr}
\noalign{\smallskip}\hline\noalign{\smallskip}
element & $\Delta \epsilon_{\rm A}$ & $\sigma$ & $[X/H]_{\rm A}$ 
& $\Delta \epsilon_B$ & $\sigma$ &  $[X/H]_B $ \\
\noalign{\smallskip}\hline\noalign{\smallskip}
C   & $+$0.10  &  0.09  &  $-$0.22  & $-$0.01  & 0.12  & $-$0.33   \\
N   & $-$0.16  &  :     &  $-$0.59  & $+$0.06  & 0.10  & $-$0.37   \\
O   & $-$0.04  &  0.04  &  $-$0.34  & $-$0.01  & 0.05  & $-$0.31   \\
Mg  & $-$0.11  &  0.07  &  $-$0.27  & $<+$0.17 & 0.10  & $-$0.09   \\
Si  & $-$0.01  &  0.10  &  $-$0.24  & $+$0.01  & 0.12  & $-$0.22   \\
\noalign{\smallskip}\hline
\end{tabular} \\
\end{table}

\begin{table}
\caption{\label{tab:abu2} Abundances [X/H] for the 
early-B stars in NGC2244 (2nd column) compared to the inner-disk early-B 
abundances (3rd column). Also, stellar abundances for CNO elements could be
compared to abundances derived for Orion Nebula
(4th column), and with solar abundances (5th column).   }
\begin{tabular}{ccccc}
\noalign{\smallskip}\hline\noalign{\smallskip}
Element &  NGC~2244$^1$  & B dwarfs$^2$   & Orion$^3$ & Sun$^4$ \\
\noalign{\smallskip}\hline\noalign{\smallskip}
C   & 8.23$\pm$0.06 &  8.27$\pm$0.12  & 8.42$\pm$0.02 & 8.41$\pm$0.03   \\
N   & 7.53$\pm$0.09 &  7.62$\pm$0.12  & 7.65$\pm$0.09 & 7.80$\pm$0.04   \\
O   & 8.56$\pm$0.03 &  8.57$\pm$0.06  & 8.51$\pm$0.03 & 8.66$\pm$0.03  \\
Mg  & 7.45$\pm$0.12 &  7.48$\pm$0.18  &  -    &  7.54$\pm$0.06  \\
Si  & 7.26$\pm$0.02 &  7.25$\pm$0.23  &  -    &   7.54$\pm$0.05    \\
Al  & 6.17$\pm$0.09 &  6.13$\pm$0.19  &  -    &   6.47$\pm$0.07   \\ 
\noalign{\smallskip}\hline
\end{tabular} \\
\smallskip

References: [1] This work and Vrancken et al.\ (\cite{paperII}); 
[2] Daflon et al.\ (\cite{ab2004}); [3] Esteban et al.\ (2004, 2005); [4]
  Daflon et al.\ \cite{ab2004} and references therein.
\end{table}

\section{\label{sec:concl} Discussion and conclusions}

We performed a detailed spectral line analysis on {\em disentangled} 
component spectra of the eclipsing early-B binary V578\,Mon in 
the open cluster NGC\,2244. Both spectra have rotationally broadened 
lines. By comparison with spectra of single stars in the same open cluster, 
temperature-dependent, faint spectral features are shown to reproduce well 
in the disentangled spectra. This validates a detailed quantitative analysis 
of such component spectra. The main concern in the error budget are the 
(coupled) undulations in the continua of disentangled spectra, which have 
to be eliminated with care. 

A differential abundance analysis relative to a sharp-lined single star,
as applied earlier in this cluster to single stars rotating faster than 
the components of V578\,Mon, revealed abundances in agreement with the 
cluster stars studied by Vrancken et al.\ (\cite{paperII}) and the large 
inner-disk sample of Daflon et al.\ (\cite{ab2004}). We conclude that 
methods applicable to observed single star spectra perform well on disentangled 
spectra given that the latter are carefully normalised to their intrinsic 
continua. Since the fundamental stellar and atmospheric parameters of 
eclipsing binaries can be known more precisely than in the case of single 
stars, this opens interesting possibilities for studying physical processes 
evolving during the main-sequence life time of the stars.

\begin{acknowledgements}

In particular, our thanks are going to Dr.~Keith Butler for his cooperation
in implementing his computer codes for NLTE analysis. Also, we are indebded
to Prof.~Philip Dufton, referee of our paper, for his constructive comments.
This research was carried out in the framework of the project `IUAP P5/36' 
financed by the Belgian Science Policy. KP acknowledges financial support 
from Croatian Ministry of Science and Technology through research grant
\#0119254. He extends his thanks to ROB for hospitality in July/August 2002 
and June/July 2003, when this work was elaborated.
\end{acknowledgements}

\end{document}